%% file: HIG-25-004_temp.tex
\documentclass[11pt,twoside,a4paper,cmspaper,final,collab]{cms-tdr}
\def\svnVersion{ccf5a69-D}\def\svnDate{2026/08/24}\def\cmsCernNoTag{CERN-EP-2026-211}\def\cmsCernDate{\today}\def\cmsMessage{Submitted to Physical Review Letters}
\begin{document}\cmsNoteHeader{HIG-25-004}

\ifthenelse{\boolean{cms@external}}{\renewcommand{\appendixname}{End Matter}}{}
\providecommand{\EndMatter}{\appendixname\xspace}

\newcommand{\GH}{\ensuremath{\Gamma_{\PH}}\xspace}
\newcommand{\GHSM}{\ensuremath{\Gamma_{\PH}^\mathrm{SM}}\xspace}

\newcommand{\ggH}{\ensuremath{\Pg\Pg\PH}\xspace}
\newcommand{\VBF}{\ensuremath{\mathrm{VBF}}\xspace}
\newcommand{\VH}{\ensuremath{\PV\PH}\xspace}

\newcommand{\sigmaI}{\ensuremath{\sigma_i}\xspace}
\newcommand{\cgG}{\ensuremath{c_{\Pg\PGg}}\xspace}
\newcommand{\GG}{\ensuremath{\PGg\PGg}\xspace}
\newcommand{\mH}{\ensuremath{m_{\PH}}\xspace}
\newcommand{\mgg}{\ensuremath{m_{\GG}}\xspace}
\newcommand{\propH}{\ensuremath{P_{\PH}}\xspace}
\newcommand{\ggGG}{\ensuremath{\Pg\Pg\to\PGg\PGg}\xspace}
\newcommand{\qgqGG}{\ensuremath{\PQq\Pg\to\PQq\PGg\PGg}\xspace}

\newcommand{\SggH}{\ensuremath{\mathcal{S}_{\ggH}}\xspace}
\newcommand{\I}{\ensuremath{\mathcal{I}}\xspace}
\newcommand{\SVBF}{\ensuremath{\mathcal{S}_{\VBF}}\xspace}
\newcommand{\SVH}{\ensuremath{\mathcal{S}_{\PV\PH}}\xspace}
\newcommand{\Bkg}{\ensuremath{\mathcal{B}}\xspace}
\newcommand{\muV}{\ensuremath{\mu_{\mathrm{V}}}\xspace}
\newcommand{\muF}{\ensuremath{\mu_{\mathrm{F}}}\xspace}

\cmsNoteHeader{HIG-25-004}
\title{Constraint on the Higgs boson width in the diphoton final state using signal-background interference in proton-proton collisions at \texorpdfstring{$\sqrt{s}=13\TeV$}{sqrt(s) = 13 TeV}}

\date{\today}

\abstract{The standard model Higgs boson, with a mass of 125\GeV, has a predicted decay width, $\GH$, of 4.1\MeV. This Letter presents the first constraint on $\GH$ obtained through the interference between the resonant $\PH \to \GG$ process and the continuum $\GG$ production, both produced via gluon-gluon fusion. The analysis is performed using proton-proton collision data at $\sqrt{s}=13\TeV$, collected by the CMS experiment between 2016 and 2018, corresponding to an integrated luminosity of 138\fbinv. The observed (expected) limit on the Higgs boson width is $\GH<92\,(138)\MeV$ at 95\% confidence level. This is the most stringent constraint on $\GH$ from on-shell Higgs boson measurements to date.}

\hypersetup{
pdfauthor={CMS Collaboration},
pdftitle={Constraint on the Higgs boson width in the diphoton final state using signal-background interference in proton-proton collisions at sqrt(s)=13 TeV},
pdfsubject={CMS},
pdfkeywords={CMS,Higgs}}

\maketitle

Within the standard model (SM) of particle physics, the Higgs boson
(\PH) mass, $\mH$, is a free parameter in the Higgs sector. For the
boson with a mass of approximately 125\GeV observed by the ATLAS and
CMS Collaborations~\cite{Aad:2012tfa,Chatrchyan:2012xdj,Chatrchyan:2013lba},
the partial decay widths to the fermions and the gauge bosons are
calculable analytically, giving rise to a total width, $\GH$, of
approximately $4.1 \MeV$~\cite{yr4}. A deviation from this
prediction could be a hint of new physics.

All measurements of \PH production cross sections depend on the total
width and on the couplings. However, a flat direction in this plane
exists, such that an increase in $\GH$ by a factor $\alpha$ due to an
undetectable decay mode cannot be disentangled from a rescaling of all
the Higgs couplings by a common factor
$1/\sqrt[4]{\alpha}$~\cite{flatdirection}. Therefore, the measurement
of $\GH$ is a key ingredient in the characterization of the Higgs
boson.
The experimental resolution limits direct measurements of $\GH$ in the
$\PH\to\PGg\PGg$ or $\PH \to \PZ\PZ\to 4\ell$ decay channels
($\ell=\Pe,\mu$). The most stringent constraint to date from direct
measurements in these channels is $\GH < 330\MeV$ at 95\% confidence
level (\CL)~\cite{masszz}.
The latest indirect constraints on $\GH$ from unitarity of the \PH
couplings by the CMS and ATLAS Collaborations are $\GH/\GHSM =
0.98^{+0.31}_{-0.25}$ and $\GH/\GHSM = 0.64 ^{+0.40}_{-0.25}$,
respectively~\cite{CMS:2018uag,ATLAS:2015egz}.
The best available precision on $\GH$ is currently obtained from
indirect measurements, by comparing on-shell and off-shell \PH
production rates for $\PH\to\PV\PV$ decays
($\PV=\PW,\PZ$)~\cite{caola,masszz,CMS:2026igg,ATLAS_offshell_width,hww}. To
date, the most precise constraint is $\GH = 3.0^{+2.0}_{-1.5}\MeV$, as
measured by the CMS Collaboration~\cite{masszz}.  This Letter presents a constraint on
$\GH$ using on-shell $\PH\to\PGg\PGg$ decays, via the interference
between \PH production via gluon-gluon fusion ($\ggH$) and the
continuum $\Pg\Pg\to\PGg\PGg$ process~\cite{martin,dixonli,liu}.

The rate of $\Pg\Pg\to \PH \to \PGg\PGg$ production is
proportional to $\cgG^2/\GH$, where $\cgG$ is the product of the
effective \PH couplings to gluons and photons, and the rate of the
continuum $\Pg\Pg\to\PGg\PGg$ production does not depend on $\GH$. 
The contribution of the interference to the total cross section,
defined as the difference between the inclusive cross section for
$\Pg\Pg$-initiated processes with a $\PGg\PGg$ final state and
the sum of the cross sections of $s$ and $b$, is termed ``interference
cross section'' ($\sigmaI$). This cross section determines the total
rate of $s+b+i$, and its value is proportional to $\cgG$. Therefore,
by relating it to the cross section of $s$, it is possible
to disentangle $\cgG$ and \GH.  

The present analysis employs proton-proton ($\Pp\Pp$) collision data
collected with the CMS detector between 2016 and 2018 at
$\sqrt{s}=13\TeV$, and corresponding to an integrated luminosity of
138\fbinv~\cite{CMS-LUM-17-003,CMS-PAS-LUM-17-004,CMS-PAS-LUM-18-002}.

The CMS apparatus~\cite{CMS:2008xjf,CMS:2023gfb} is a multipurpose,
nearly hermetic detector, designed to trigger
on~\cite{CMS:2020cmk,CMS:2016ngn,CMS:2024aqx} and identify electrons,
muons, photons, and (charged and neutral)
hadrons~\cite{CMS:2020uim,CMS:2018rym,CMS:2014pgm}. The global
``particle-flow'' algorithm~\cite{CMS:2017yfk} aims to reconstruct all
individual particles in an event, combining information provided by
the all-silicon tracker, and by the crystal electromagnetic (ECAL) and
brass-scintillator hadron calorimeters, operating inside the
3.8\unit{T} superconducting solenoid, with data from the
gas-ionization muon detectors embedded between the layers of the
flux-return yoke outside the solenoid. The reconstructed particles are
used to build \PGt leptons, jets, and missing transverse
momentum~\cite{CMS:2018jrd,CMS:2016lmd,CMS:2019ctu}.

In this analysis, leading order (LO) and next-to-leading
order (NLO) amplitudes in quantum chromodynamics (QCD) are taken into
account for gluon-gluon fusion processes, together
with the LO contribution of the tree-level process
\qgqGG~\cite{dixonli}. Samples of \PH events produced via \ggH, vector boson fusion (\VBF),
and vector boson associated production (\VH, $\PV=\PW,\PZ$) are
simulated with the \MGvATNLO 2.4.2 matrix-element
generator~\cite{MadGraph} at NLO.  These parton-level samples are
interfaced with \PYTHIA version 8.230 \cite{Sjostrand:2014zea} for
parton showering and hadronization, with the \textsc{CP5}
tune~\cite{cp5}.  Events produced via \ggH are weighted as a function
of the transverse momentum (\pt) of the Higgs boson and the number of
jets in the event, to match the prediction from the NNLOPS program
[43], as in Ref.~\cite{paper-stxs}. The parton distribution functions
are taken from the \textsc{NNPDF} 3.1 set~\cite{NNPDF:2017mvq} at
next-to-NLO (NNLO).  Other signal processes, such as \PH production in
association with a top quark or with a top quark-antiquark pair, are
not taken into account in this analysis, for reasons discussed below.

Simulated samples that describe the interference between the \ggH
signal and the \ggGG or \qgqGG background processes are produced at
NLO in QCD using the \SHERPA~2.2.11 generator~\cite{sherpa}, for \mH
values of 120, 125, and 130\GeV, and $\GH=4.1 \MeV$. Additional
samples were produced at 125\GeV \mH and $\GH/\GHSM =
1,\,2,\,\ldots,\,5,\,10,\,20,\,50$.  All the amplitudes considered in Ref.~\cite{dixonli} are included for this analysis. Hadronization and
parton showering are described using the Catani--Seymour dipole
factorization~\cite{Catani:1996vz}.  The NNLO effects on the
interference were calculated recently~\cite{devoto}, yielding a larger
inclusive interference cross section with respect to \SHERPA. No
comparison with differential cross sections predicted by \SHERPA is
available, nor Monte Carlo generator is publicly available. Therefore
the latter calculation is not included in this work.

The dominant background processes are continuum diphoton production,
$\PGg$+jets, and multijet production, with one or more jets
misidentified as a photon. These backgrounds are modeled using data
sidebands, excluding the region in diphoton invariant mass (\mgg)
between 115 and 135\GeV.

All simulated samples are propagated through the full CMS detector
simulation using the \GEANTfour package~\cite{g4}. The samples include
multiple nearly simultaneous $\Pp\Pp$ collisions happening in the same
or adjacent bunch crossing, termed pileup.  The simulated events
are reweighted to reproduce the pileup distribution observed in data.

In the narrow-width approximation, the differential production cross
section as a function of \mgg for the \ggH signal and the interference
components can be expressed as
\begin{align}
    \dd{\sigma_s}{\mgg} &= \frac{\cgG^2 S}{(\mgg^2 - \mH^2)^2 + \mH^2 \GH^2},\label{eq:sigmaS}\\
    \dd{\sigma_i}{\mgg} &= \cgG \frac{(\mgg^2 - \mH^2) A_R - \mH \GH A_I}{(\mgg^2 - \mH^2)^2 + \mH^2 \GH^2},
    \label{eq:sigmaI}
\end{align}
where $S$, $A_R$, and $A_I$ are numerical coefficients independent of
the \PH couplings and width \cite{dixonli},
detailed in \EndMatter~\ref{app:crosssections}.
 
\begin{figure}[htbp!]
    \centering
    \includegraphics[width=\linewidth]{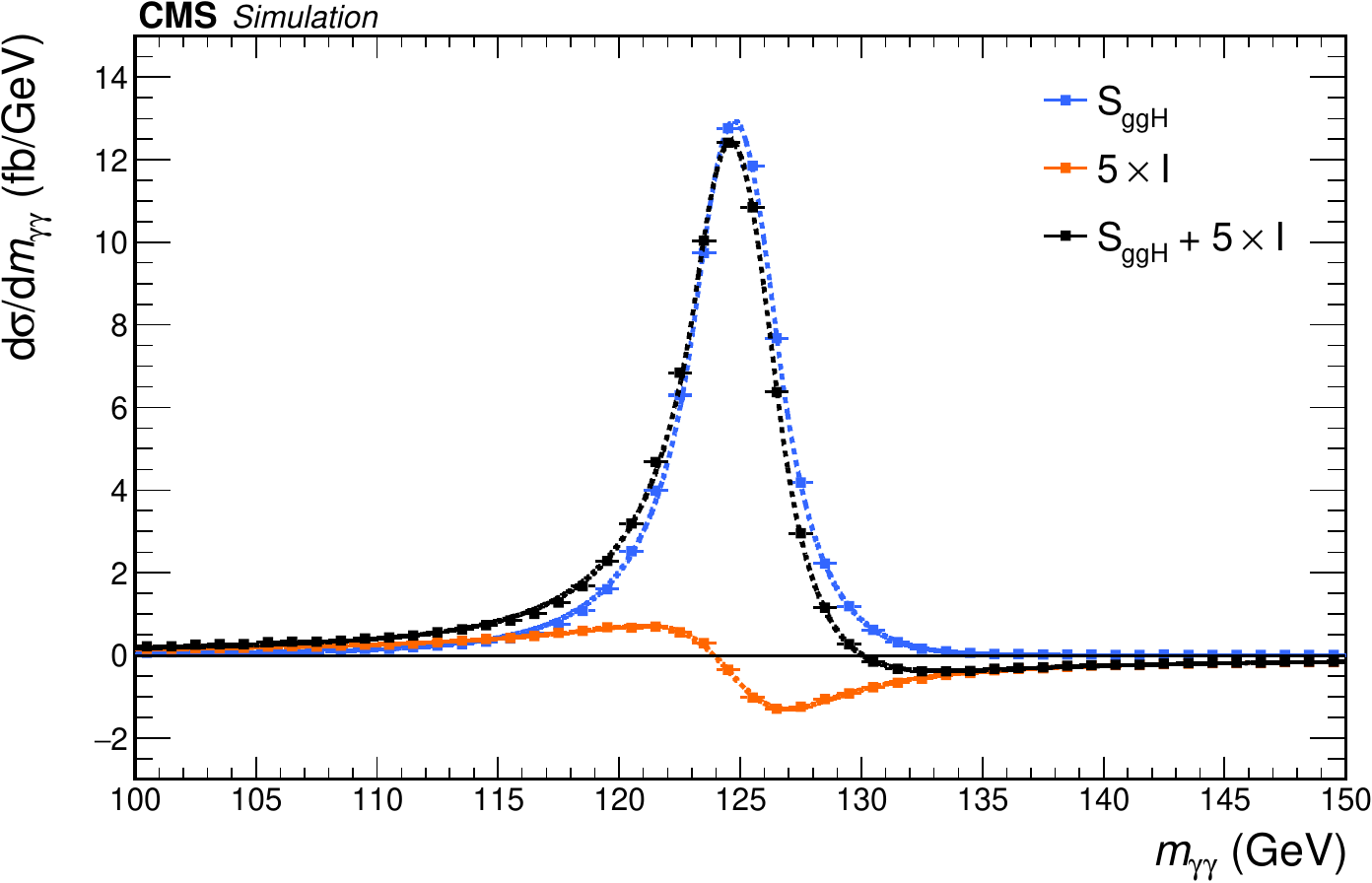}

    \caption{ Diphoton invariant mass spectrum after full detector simulation, for the $\ggH$ signal
        component of Eq.~\eqref{eq:sigmaS}, denoted \SggH (blue points), for the interference
        contribution of Eq.~\eqref{eq:sigmaI}, denoted \I and amplified by a factor 5
        for better visualization (orange points), and their sum (black points), for samples with $\mH=125\GeV$ and
        $\GH=\GHSM$.  \label{fig:spectrumint}}
\end{figure}

The production cross section of the pure \ggH signal process, obtained
by integrating Eq.~\eqref{eq:sigmaS}, is
proportional to $\cgG^2/\GH$. By inverting this relation, the relationship between the effective \PH
couplings and \GH can be parameterized as~$\cgG=\sqrt{\smash[b]{\muF\GH/\GHSM}}$. In this expression,
the signal strength $\muF$ is the measured rate of events of the \ggH
component with respect to its SM value, and \GHSM is the total width
expected for the SM Higgs boson with $\mH=125\GeV$. The differential cross section for the interference ($i$) in
Eq.~\eqref{eq:sigmaI} has an asymmetric lineshape with a negative
contribution above the Higgs boson pole mass, as shown in
Fig.~\ref{fig:spectrumint}.

Since the $A_R$ factor in
Eq.~\eqref{eq:sigmaI} changes sign from positive to negative when $\mgg=\mH$, with this contribution, the overall
$\mgg$ distribution deviates notably from the `naive' Breit--Wigner function. 
This modification, which can be parameterized
as a function of \GH, provides the primary handle for measuring \GH when \PH is produced on shell.

The differential cross sections of $s$ and $i$ can be expressed
explicitly in terms of the $\ggH$ signal strength $\muF$, the
coupling modifier $\cgG$, and $\GH$:
\begin{equation}
  \dd{\sigma_s}{\mgg} = \frac{\cgG^2}{\GH} \GHSM \SggH^\mathrm{SM} = \muF \SggH, \dd{\sigma_i}{\mgg} = \cgG \I^\mathrm{SM},
\end{equation}
where $\SggH^\mathrm{SM}$ and $\I^\mathrm{SM}$ represent the
differential cross sections expected in the SM for the $\ggH$ signal and interference
components in $\mgg$. These functions do not depend on \cgG or \GH
in the limit of a resonance width that is much smaller than the
experimental resolution of $\mathcal{O}(1)\GeV$. This means that
templates obtained for the simulation of an SM \PH can be used in the
range of \GH considered in the analysis with negligible impact on the
results, as verified by using dedicated simulated samples.

A convenient linear combination of the $s$ and $i$ differential
cross sections is:
\begin{equation}
    \dd{\sigma_s}{\mgg} + \dd{\sigma_i}{\mgg} =
    (\muF-\cgG)\SggH^\mathrm{SM} + \cgG(\SggH+\I)^\mathrm{SM},
\end{equation}
which can be described as a function of two templates, one for
$\SggH^\mathrm{SM}$ and one for $(\SggH+\I)^\mathrm{SM}$, both
positive definite in a region of approximately 4\GeV around the \PH
peak. This is needed to employ probability density functions (pdfs) in
the statistical modeling. Conversely, the $\I^\mathrm{SM}$ component
is always negative for $\mgg>\mH$, and therefore cannot be interpreted
as a pdf.
The total diphoton production cross section, including the \ggH
process $s$, the \VBF and \VH productions (\PV), and the continuum
background $b$, can be written as a function of \GH
as:
\begin{multline}
 \dd{\sigma_s}{\mgg} + \dd{\sigma_i}{\mgg} +
 \dd{\sigma_s^{\PV}}{\mgg} + \dd{\sigma_b}{\mgg} =  \\
\left(\muF - \sqrt{\muF \frac{\GH}{\GHSM}} \right)\SggH^\mathrm{SM} +
\sqrt{\muF \frac{\GH}{\GHSM}} (\SggH + \I)^\mathrm{SM} \\ 
+ \muV
 \left(\SVBF^\mathrm{SM} + \SVH^\mathrm{SM}\right) + \Bkg,
\label{eq:mix}
\end{multline}
where $\muV$ represents the signal strength in the
$\VBF$ and $\VH$ processes and \Bkg represents the background component. 
The minor and non-consequential contribution from the 
interference induced by the quark-gluon initiated processes is
included in the gluon-gluon fusion interference templates $(\SggH+\I)^\mathrm{SM}$.

Events used in this analysis are selected using a diphoton trigger with asymmetric photon
\pt thresholds of 30 and 18 \GeV in 2016, and 30 and 22 in 2017 and 2018
data, with a further selection based on calorimeter
variables and isolation to reject jets and tracks~\cite{paper-stxs}.

Photon candidates are reconstructed and selected as described in Ref.~\cite{paper-stxs}, 
including requirements on kinematic properties of photons and ECAL shower-shape variables, isolation with respect to other 
surrounding reconstructed particles, a veto on electrons and on the energy deposited in the hadronic calorimeter behind the 
photon ECAL cluster. A boosted decision tree (BDT) algorithm, labeled diphoton BDT in what follows, is trained to discriminate $\PH\to\PGg\PGg$ events from diphoton background processes~\cite{paper-stxs}. 
This BDT employs photon identification variables and kinematic properties as inputs. 
The training minimizes correlations between the BDT output and the diphoton invariant mass, avoiding sculpting of the
$\mgg$ distribution. Jets are reconstructed and identified as described 
in Ref.~\cite{paper-stxs}.

Selected diphoton candidates are categorized according to their kinematic properties, to define regions 
with different sensitivity to the interference effect and different signal purity. The first step loosely 
targets \VBF and \ggH events with two associated jets, which have a smaller or opposite interference 
contribution~\cite{yr4}, assigned to a dedicated two-jet category, leaving a sample enriched in events with at most one jet for the remaining categories. This two-jet category is defined by requiring at least two reconstructed jets, with leading (subleading) $\pt>30$ $(20)\GeV$ and dijet invariant mass greater than $100\GeV$.

Events without a dijet pair satisfying the requirements described above
are further divided into five regions that target \ggH production. This categorization is based on the \pt of the
diphoton candidate, as the $(\SggH + \I)$ lineshape depends on this
variable. The dependence is the result of the underlying variation of
the signal and interference cross sections as a function of the
diphoton $\pt$, as detailed in Ref.~\cite{dixonli}, where this effect
is studied using the shift in the $\mgg$ peak position.

The five regions targeting \ggH production are each divided in two,
depending on the diphoton BDT output, to increase the purity in signal
events for one of the two categories.  The diphoton BDT category
boundaries depend on the data-taking period, and were chosen by
minimizing the ratio between the
expected $\mgg$ resolution and the statistical significance of the
signal, summed over the categories. The optimized $\pt$ boundaries are: $[15,\, 21,\, 31,\, 47,\,
  77,\, +\infty]\GeV$.  In total 11 categories are employed:
one targeting two-jet events, and 10 categories for the remaining
events.

To extract \GH, an extended binned maximum likelihood fit~\cite{lf} to
the $\mgg$ spectrum is performed simultaneously in all 11 analysis
categories.
The signal models are derived for each combination of category and
production process, with separate pdfs built from simulation for four
possible production mechanisms: $\SggH^\mathrm{SM}$, $(\SggH +
\I)^\mathrm{SM}$, $\SVBF^\mathrm{SM}$, and $\SVH^\mathrm{SM}$. 

 To build the signal model, each of these pdfs is multiplied by the expected signal yield evaluated from simulation. Each pdf, 
 together with its expected rate, is treated as a process contributing to the total signal.

Other signal processes, such as \PH production in association with top
quarks, are not explicitly taken into account, because they have a
negligible contribution to the signal yield in all the categories
targeting events with fewer than two jets.  The signal models are
derived by fitting the simulated $\mgg$ spectra for the four processes
with a double-sided Crystal Ball (DCB) function~\cite{dcb,geiser}, a
pdf characterized by a Gaussian core and two power law tails, with parameters dependent on $\mH$.
Further details on the procedure employed to derive the signal model and the
interpolation with respect to $\mH$ can be found in
Ref.~\cite{mass2016}.

The $(\SggH+\I)$ parameterization is obtained by fitting the range
118--128\GeV\ and extrapolating the fitted function elsewhere.
The fit range is restricted to the region where the interference
contribution is most relevant and the parameterization provides a
stable description of the signal shape.
This approximation is valid because for $\mgg\gtrsim128\GeV$ the
signal-to-background ratio is very small, and the effect of the
negative component of $\I$, shown in Fig.~\ref{fig:spectrumint}, is covered
by the statistical uncertainty in the background. This procedure has been
validated with pseudo-experiments by generating data sets
with the sum of the $(\SggH+\I)$ and $\Bkg$ templates, including the
negative tail, and fitting them with the approximated $(\SggH+\I)$ model.
The induced bias on the results is found to be negligible.

The nonresonant backgrounds are fitted with empirical analytical
models derived from $\mgg$ data sidebands in the ranges
100--115\GeV and 135--180\GeV. Families of monotonically decreasing
functions, described in Ref.~\cite{paper-stxs}, are tested for each
category, and in the signal extraction fit the choice of the function
is included in the total uncertainty by treating it as a discrete
nuisance parameter~\cite{discr}.

The total $\mgg$ pdf, for each category $c$, is modeled as the sum of
the pdfs for each of the four signal processes
($\SggH,\SggH+\I,\SVBF,\SVH$), normalized with their simulated event yields, and weighted with
$(\muF-\sqrt{\smash[b]{\muF \GH/\GHSM}})$
and $\sqrt{\smash[b]{\muF\GH/\GHSM}}$ for the $\SggH$ and $(\SggH+\I)$
processes, respectively, while a single parameter $\muV$, scales both the \VBF and \VH pdfs.

The likelihood function, used for the fit, is then defined as a
product of Poissonian conditional probabilities over all categories,
with $\GH$ as the parameter of interest of the fit, while the parameter
$\mH$, which regulates the position of the Higgs boson peak in the
signal pdf, as well as the two signal strengths, $\muF$ and $\muV$,
are left floating in the fit.  To avoid the discontinuity in the
derivative of the likelihood due to its dependence on $\sqrt{\smash[b]{\GH}}$,
the minimization is performed with respect to a transformed variable,
$\lambda=\sqrt{\smash[b]{\GH/\GHSM}}$. The constraint on $\lambda$ is then
converted to a constraint on \GH.  More details on the event yield
parameterization and on the likelihood are
provided in \EndMatter~\ref{app:likelihoods}.

Uncertainties that alter the shape of the signal $\mgg$ distributions
are incorporated into the analytical templates as nuisance parameters,
by allowing the DCB parameters to be modified with Gaussian constraints. Uncertainties that
affect the signal normalization are described with log-normal
pdfs. The constraint on each nuisance parameter is modeled via an
appropriate pdf~\cite{combine}.

\begin{figure}[htbp!]
\centering
\includegraphics[width=\linewidth]{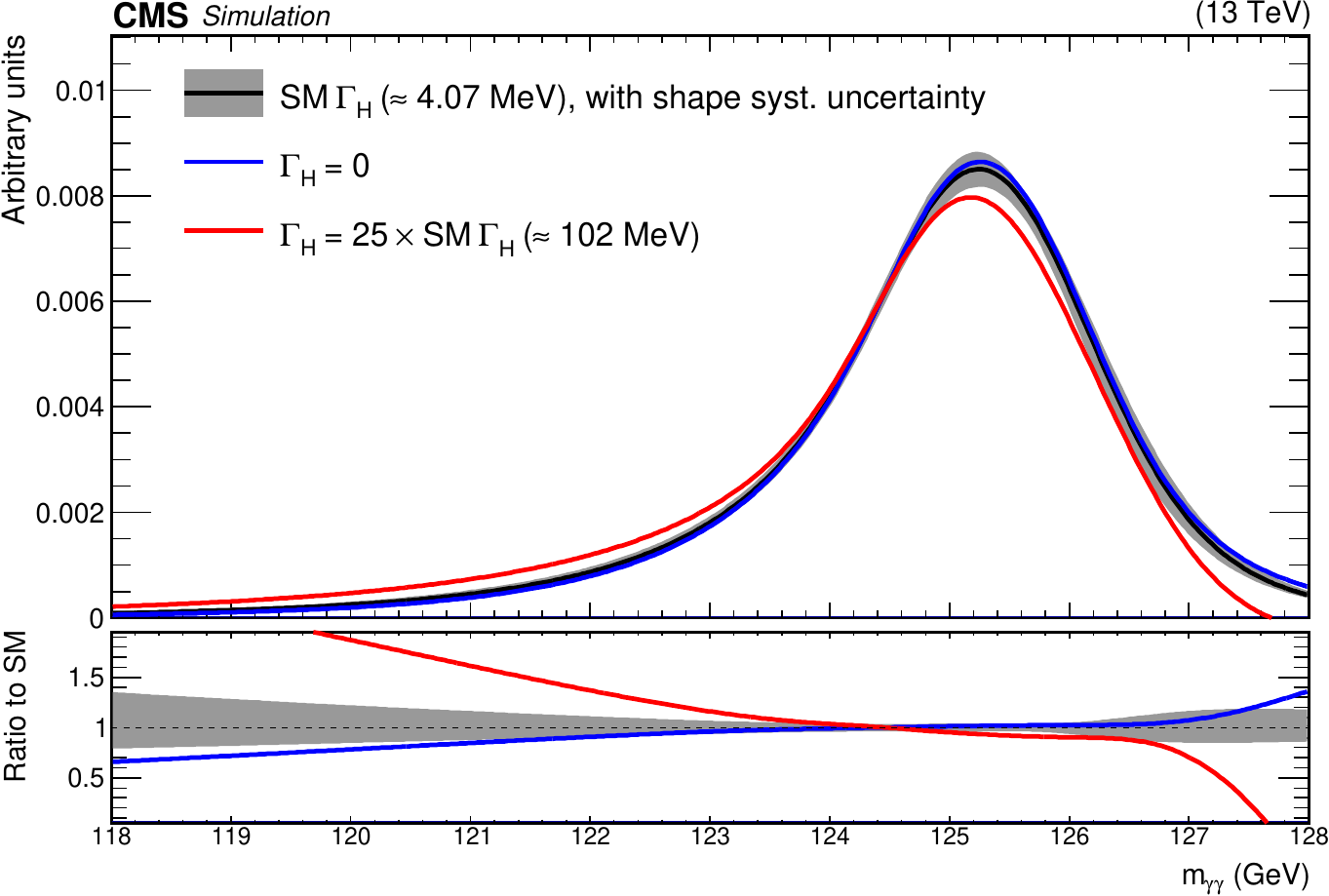}
\caption{
The nominal signal shape in $\mgg$ for $\GH=\GHSM$, shown with the 68\% \CL envelope
of all the experimental shape systematic uncertainties (grey band), together
with the $\GH = 0$ and $25\,\GHSM$ hypotheses (blue and red lines,
respectively). The shown shapes are obtained as the weighted sum of each process
and category. The weight for each category is proportional to
$S/(S+B)$, where $S$ and $B$ are, respectively, the expected signal and
background yields in a narrow window around the $\PH\to\PGg\PGg$ mass peak.
\label{fig:systs}}
\end{figure}

The largest experimental systematic uncertainties are photon
resolution uncertainties for photons with $\pt>50\GeV$, with maximum
values of 0.8\%, and uncertainties in the modeling of the material
budget in front of the ECAL. Those effects could mimic the effect of the interference on the
\PH lineshape. A possible nonlinearity in the photon
energy scale, which directly affects the observed peak position, is
parameterized as a Gaussian nuisance parameter with a standard
deviation of 0.2\%~\cite{paper-stxs}, fully correlated across
different categories, but uncorrelated between data-taking years.
The pdfs for the SM signal model and for the beyond-the-SM
hypotheses $\GH=0$ and $\GH=25\,\GHSM$, together with the envelope of
all experimental systematic uncertainties in shape, are shown in
Fig.~\ref{fig:systs}.  The value $\GH=25\,\GHSM$ is chosen because it is
close to the expected 95\% \CL limit of the analysis.

The most important theoretical uncertainties considered in the
analysis are the renormalization ($\mu_R$) and factorization ($\muF$) scales, and 
the strong coupling constant \alpS. Their effect is included in
the simulated signal samples following the guidelines in
Ref.~\cite{yr4}. For the interference cross section, a global 7\%
normalization uncertainty is applied, estimated with \SHERPA to generate alternative samples corresponding to each
parameter variation ($\alpS, \mu_R, \muF$). This uncertainty is incorporated by considering an alternative pdf for $\SggH+\I$.

\begin{figure}[htbp!]
\centering
\includegraphics[width=\linewidth]{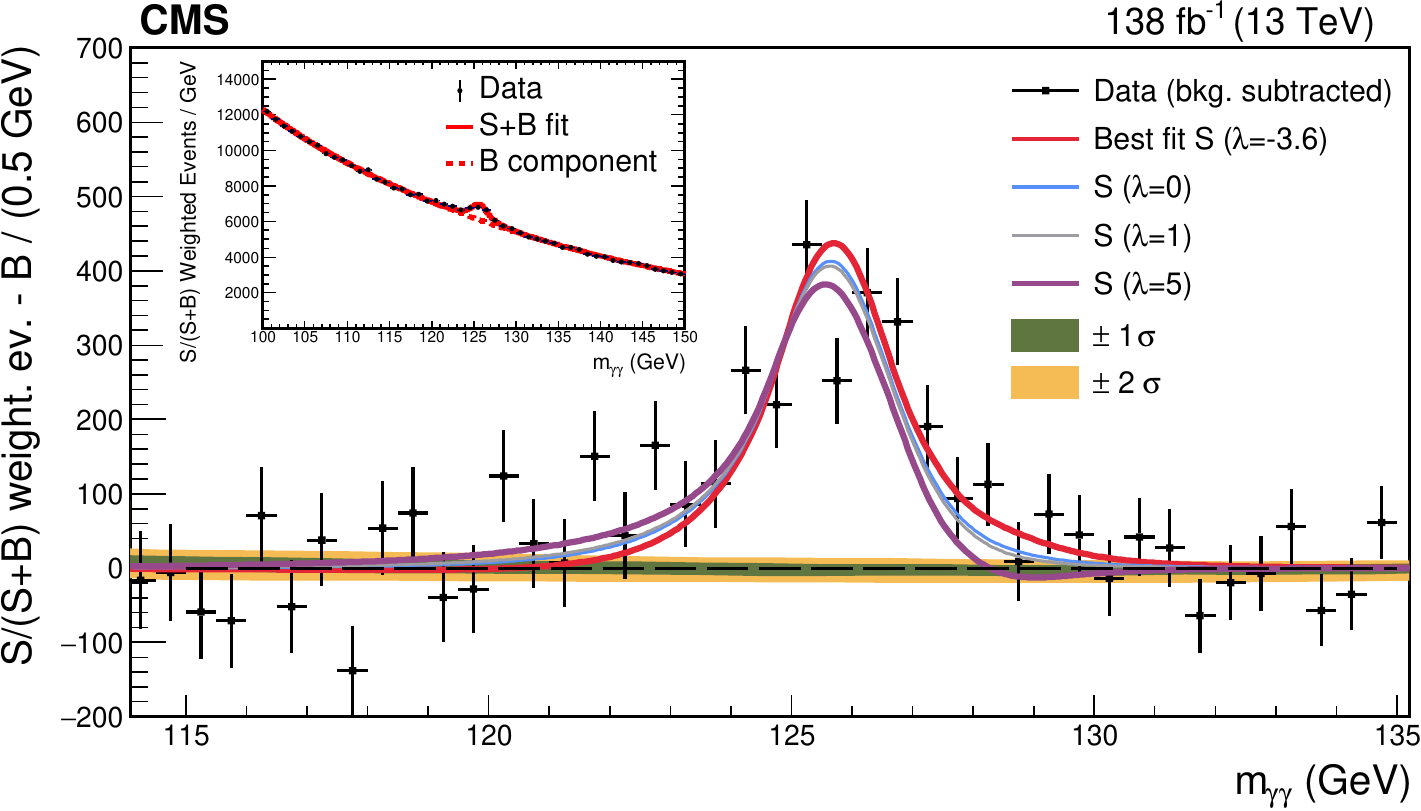}
\caption{ Weighted $\mgg$  distribution in data, after
  background subtraction, with weights proportional to $S/(S+B)$, in a
  zoomed range around the observed \mH. The best fit signal model (red
  line) is superimposed on the data, together with signal predictions
  in which all parameters except $\lambda = \sqrt{\smash[b]{\GH/\GHSM}}$ are set to their best fit
  values. The parameter $\lambda$ is evaluated at several different
  values of $\lambda=0,\,1,\,5$ (blue, gray, purple lines). In the
  inset plot, the data and fitted distributions are shown before
  background subtraction in the full fit range.
\label{fig:fit}}
\end{figure}

The results are extracted by performing
a simultaneous binned fit to the $\mgg$ distribution in data in all
categories and data-taking periods, considering the range
$100<\mgg<180\GeV$. The CMS statistical analysis tool \textsc{Combine}~\cite{combine}
is used to perform this fit. A goodness-of-fit check yields a \textit{p}-value
greater than 90\%. The background-subtracted observed data, together
with the signal plus background model at the best fit point, are shown
in a restricted $\mgg$ range in Fig.~\ref{fig:fit}. Signal plus
background models corresponding to alternative values of $\lambda$ are
also shown.

The measured value of $\lambda$ is found to be $\lambda = -3.6^{+2.5}_{-2.4}$, which is compatible with the SM expectation of $\lambda=1$ within two standard deviations. The values of the signal strengths $\muF$ and $\muV$ are found to be
consistent with the SM expectation. 

The Feldman--Cousins approach~\cite{Feldman:1997qc} is used to provide
confidence intervals on $\GH$, by incorporating the physical boundary
$\GH>0$, and generating $10^3$ pseudo-experiments for each tested
value of $\GH$. The test statistic employed to derive the limits is
the profile likelihood ratio~\cite{combine}. More details are
provided in \EndMatter~\ref{app:feldmancousins}.

\begin{figure}[htbp!]
\centering
    \includegraphics[width=\linewidth]{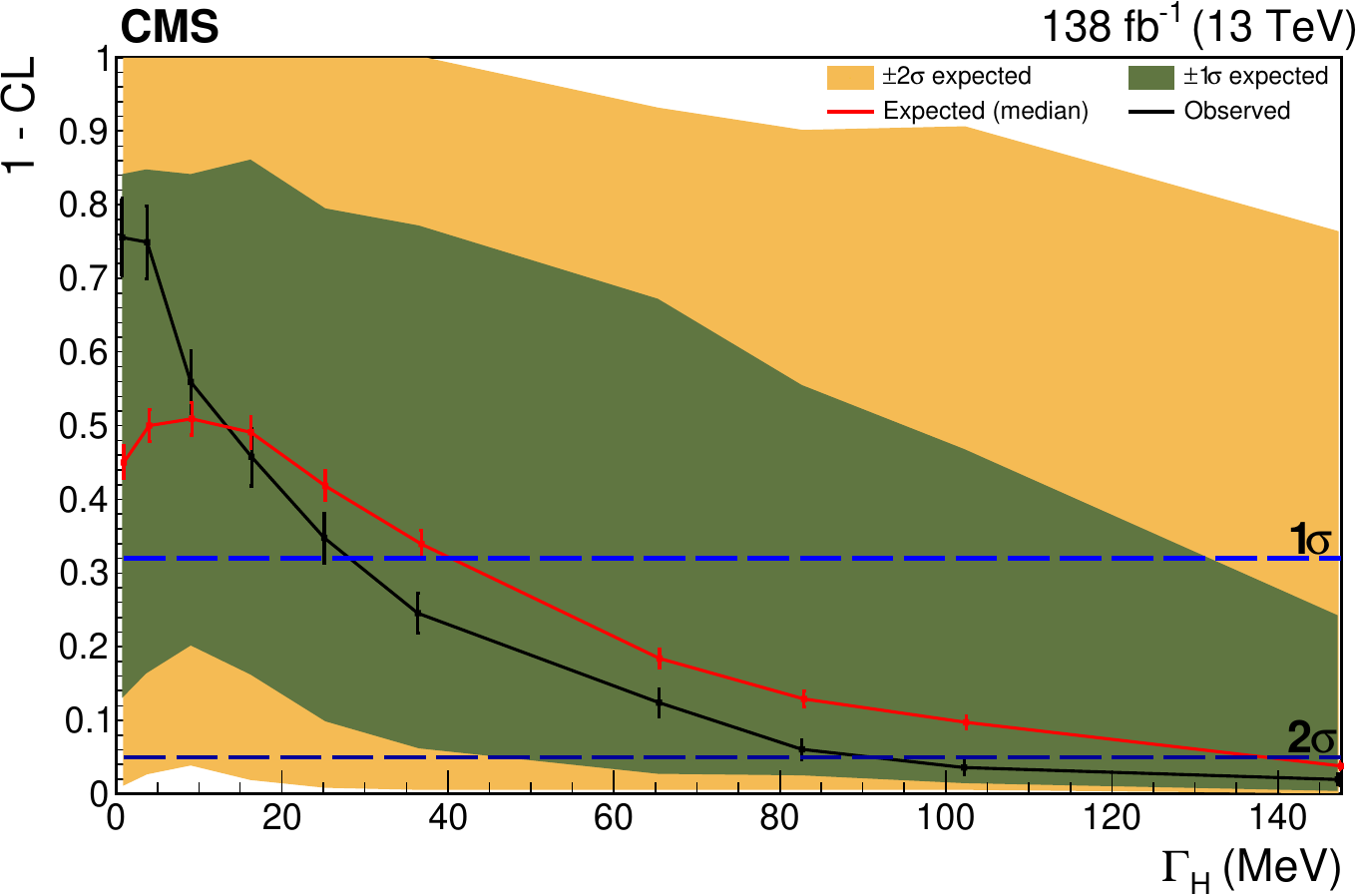}
\caption{Distribution of $1 - \CL$ as a function of $\GH$, constructed using the Feldman--Cousins approach as defined in the text.
The vertical bars on the points represent the statistical uncertainty in $1 - \CL$ due to the limited number of pseudo-experiments.
The dashed horizontal lines indicate the 68\% and 95\% confidence levels.
\label{fig:results}}
\end{figure}

The observed (expected) limit at 95\% \CL on the \PH total width,
obtained by converting the limit on $\lambda$, is $\GH<92\, (138)\MeV$, as shown in Fig.~\ref{fig:results}.

An additional approach based on the interference-induced displacement
of the \PH mass peak, proportional to the effective coupling $\cgG$,
is also studied.  Differently from the method previously described,
the $\mgg$ peak position, for $\muF=1$, for each analysis category
$k$, is parameterized as $M^{\SggH+\I}_k = M^{\SggH}_k + \alpha_k
\sqrt{\smash[b]{\GH/\GHSM}}$
where the coefficient $\alpha_k$ scales the size of the mass
displacement. The parameters $\alpha_k$ are determined from simulation
by performing a simultaneous fit to the $\mgg$ distributions
constructed with and without interference, using the same parametric
signal model. However, the dependence of the
signal lineshape on \GH is parameterized with a
Voigtian function, with the mean of its Breit--Wigner component
defined as $M^{\SggH+\I}_k$.  
In this formulation, \GH enters the likelihood fit exclusively through
the modification of the peak position for the $\SggH$
component. Although this strategy provides some sensitivity to $\GH$,
it is intrinsically limited by the achievable precision on the
determination of $\mH$ and by residual modeling uncertainties. With
this approach, an expected limit of $\GH<650\MeV$ at 95\% \CL is
obtained.  Tabulated results are provided in the HEPData record for
this analysis~\cite{hepdata}.

In summary, a constraint on the Higgs boson width $\GH$ is presented,
using for the first time the interference between the resonant
$\PH \to \GG$ and continuum $\GG$ processes, both
produced via gluon-gluon fusion. The observed (expected) limit is
$\GH<92\, (138)\MeV$ at 95\% confidence level.  This is the
most stringent constraint on $\GH$ from on-shell Higgs boson
measurements to date.

\begin{acknowledgments}
We congratulate our colleagues in the CERN accelerator departments for the excellent performance of the LHC and thank the technical and administrative staffs at CERN and at other CMS institutes for their contributions to the success of the CMS effort. In addition, we gratefully acknowledge the computing centers and personnel of the Worldwide LHC Computing Grid and other centers for delivering so effectively the computing infrastructure essential to our analyses. Finally, we acknowledge the enduring support for the construction and operation of the LHC, the CMS detector, and the supporting computing infrastructure provided by the following funding agencies: SC (Armenia), BMFWF and FWF (Austria); FNRS and FWO (Belgium); CNPq, CAPES, FAPERJ, FAPERGS, and FAPESP (Brazil); MES and BNSF (Bulgaria); CERN; CAS, MoST, and NSFC (China); MINCIENCIAS (Colombia); MSES and CSF (Croatia); RIF (Cyprus); SENESCYT (Ecuador); ERC PRG and PSG, TARISTU24-TK10 and MoER TK202 (Estonia); Academy of Finland, MEC, and HIP (Finland); CEA and CNRS/IN2P3 (France); SRNSF (Georgia); BMFTR, DFG, and HGF (Germany); GSRI (Greece); MATE and NKFIH (Hungary); DAE and DST (India); IPM (Iran); SFI (Ireland); INFN (Italy); MSIT and NRF (Republic of Korea); MES (Latvia); LMTLT (Lithuania); MOE and UM (Malaysia); BUAP, CINVESTAV, CONACYT, LNS, SEP, and UASLP-FAI (Mexico); MOS (Montenegro); MBIE (New Zealand); PAEC (Pakistan); MSHE, NSC, and NAWA (Poland); FCT (Portugal); MESTD (Serbia); MICIU/AEI and PCTI (Spain); MOSTR (Sri Lanka); Swiss Funding Agencies (Switzerland); MST (Taipei); MHESI (Thailand); TUBITAK and TENMAK (T\"{u}rkiye); NASU (Ukraine); STFC (United Kingdom); DOE and NSF (USA).
\end{acknowledgments}\section*{Data availability} Release and preservation of data used by the CMS Collaboration as the basis for publications is guided by the  \href{https://doi.org/10.7483/OPENDATA.CMS.1BNU.8V1W}{CMS data preservation, re-use and open access policy}.

\bibliography{auto_generated} 

\appendix

\section{Signal and interference cross sections}\label{app:crosssections}
In the narrow-width approximation, the differential cross
sections, as a function of \mgg, for the \ggH signal and the interference
components can be expressed as
\begin{equation*}
\begin{aligned}
    \dd{\sigma_s}{\mgg} &= \frac{\cgG^2 S}{(\mgg^2 - \mH^2)^2 + \mH^2 \GH^2},\\
    \dd{\sigma_i}{\mgg} &= \cgG \frac{(\mgg^2 - \mH^2) A_R - \mH \GH A_I}{(\mgg^2 - \mH^2)^2 + \mH^2 \GH^2},
\end{aligned}
\end{equation*}
where $S$, $A_R$, and $A_I$ are numerical coefficients that are
independent of the Higgs boson couplings and width.  The coefficient
$S$ is positive and relates the signal production cross section to the
relativistic Breit--Wigner function, which is the squared modulus of
the Higgs boson propagator $\propH(\mgg^2) = 1/(\mgg^2 - \mH^2 + i \mH
\GH)$.  The $A_R$ and $A_I$ coefficients, which are, respectively,
negative and positive, relate the interference production cross
section to the real and imaginary parts of $\propH(\mgg^2)$, by
writing $\dd{\sigma_i}{\mgg}$ as $\cgG \{A_R\text{Re}[\propH(\mgg^2)]
+ A_I \text{Im}[\propH(\mgg^2)]\}$.  The real part of this
differential cross section, labeled R-term in the following, can be
written as $(\mgg^2- \mH^2) \abs{\propH(\mgg^2)}^2$, which changes
sign at $\mgg=\mH$. The imaginary part is proportional to
$-\abs{\propH(\mgg^2)}^2$, therefore it is a negative relativistic
Breit--Wigner function~\cite{dixonli,martin}.

\section{Event yields and likelihood}\label{app:likelihoods}
The total $\mgg$ pdf, for each category $c$, is modeled as the sum of
the pdfs for each of the four signal processes
($\SggH,\SggH+\I,\SVBF,\SVH$), indexed by $j$, generically denoted
$P^s_{j,c}$, and the one for the background ($P_c^b$).  Each signal
pdf is multiplied by the product of a $\mu_j$ coefficient and the
expected event yield $n_{j,c}$, such that the yield for bin $k$ in
category $c$ is:
\ifthenelse{\boolean{cms@external}}{
\begin{multline*}
    N_c(k, \GH/\GHSM, \vec{\mu}, \mH, \vec{\theta}) =  \\
    \sum_{j}^{N_\mathrm{proc}} \mu_j(\GH/\GHSM) n_{j,c} P_{j,c}^s\left(\mgg^k, \mH ;
    \vec{\theta}\right) + P_{c}^b\left(\mgg^k; \vec{\theta}\right), 
\end{multline*}}
{
\begin{equation*}
    N_c(k, \GH/\GHSM, \vec{\mu}, \mH, \vec{\theta}) =  
    \sum_{j}^{N_\mathrm{proc}} \mu_j(\GH/\GHSM) n_{j,c} P_{j,c}^s\left(\mgg^k, \mH ;
    \vec{\theta}\right) + P_{c}^b\left(\mgg^k; \vec{\theta}\right), 
\end{equation*}
}
where $\vec{\theta}$ represents the set of constrained nuisance
parameters, which encode the systematic uncertainties. The
coefficients $\mu_j$ are defined as $(\muF-\sqrt{\smash[b]{\muF \GH/\GHSM}})$
and $\sqrt{\smash[b]{\muF\GH/\GHSM}}$ for the $\SggH$ and $(\SggH+\I)$
processes, respectively. A single parameter $\muV$, independent of
\GH, scales both the \VBF and \VH pdfs.

The likelihood function is then defined as a product of conditional probabilities
over all categories:
\ifthenelse{\boolean{cms@external}}{
\begin{multline*}
\mathcal{L}\left(\text{data} \mid \GH/\GHSM, \vec{\mu}, \mH, \vec{\theta}_s, \vec{\theta}_b\right) = \\
\prod_{c=1}^{N_\text{cat}} \prod_{i=1}^{N_{\text{bins}}} \text{Poisson}\left[n^{\text{data}}_{c,i} \mid
N_c\left(i,\GH/\GHSM, \vec{\mu}, m_\PH, \vec{\theta}\right) \right],
\label{eq:likelihood}
\end{multline*}}
{
\begin{equation*}
\mathcal{L}\left(\text{data} \mid \GH/\GHSM, \vec{\mu}, \mH, \vec{\theta}_s, \vec{\theta}_b\right) = 
\prod_{c=1}^{N_\text{cat}} \prod_{i=1}^{N_{\text{bins}}} \text{Poisson}\left[n^{\text{data}}_{c,i} \mid
N_c\left(i,\GH/\GHSM, \vec{\mu}, m_\PH, \vec{\theta}\right) \right],
\label{eq:likelihood}
\end{equation*}}
where $n_{c,i}^{\text{data}}$ is the number of observed data
events in bin $i$ of category $c$.

\section{Feldman-Cousins procedure}\label{app:feldmancousins}
The Feldman--Cousins approach~\cite{Feldman:1997qc} is used
to provide confidence intervals on $\GH$, by incorporating the
physical boundary $\GH>0$, and generating $10^3$ pseudo-experiments
for each tested value of $\GH$. The test statistic employed to derive
the limits is the profile likelihood ratio~\cite{combine}.
For each
value of $\GH$, $1 - \CL$ is defined as the probability, computed
using pseudo-experiments under the hypothesis of $\GH$ as the true
value, to obtain a value of the test statistic at least as large as a
reference value.  The observed confidence levels are obtained by using
as reference, for each value of $\GH$, the value of the test statistic
evaluated on data, while the median expected ones employ as references
the median values of the test statistic from pseudo-experiments
generated at $\GH = \GHSM$.  The edges of the 68\% and 95\% CL bands
for the expected limits are obtained by using, respectively, the
[0.16--0.84] and [0.025--0.975] quantiles of the test statistic
distribution of the pseudo-experiments at $\GH = \GHSM$.

\cleardoublepage \section{The CMS Collaboration \label{app:collab}}\begin{sloppypar}\hyphenpenalty=5000\widowpenalty=500\clubpenalty=5000\input{HIG-25-004-public-authorlist.tex}\end{sloppypar}
\end{document}

%% file: HIG-25-004-public-authorlist.tex
\cmsinstitute{Yerevan Physics Institute, Yerevan, Armenia}
{\tolerance=6000
A.~Gevorgyan\cmsorcid{0000-0003-2751-9489}, A.~Hayrapetyan, V.~Makarenko\cmsorcid{0000-0002-8406-8605}, A.~Tumasyan\cmsAuthorMark{1}\cmsorcid{0009-0000-0684-6742}
\par}
\cmsinstitute{Marietta Blau Institute for Particle Physics, Vienna, Austria}
{\tolerance=6000
W.~Adam\cmsorcid{0000-0001-9099-4341}, L.~Benato\cmsorcid{0000-0001-5135-7489}, T.~Bergauer\cmsorcid{0000-0002-5786-0293}, M.~Dragicevic\cmsorcid{0000-0003-1967-6783}, S.~Gundacker\cmsorcid{0000-0003-2087-3266}, A.K.~Guven\cmsorcid{0009-0004-5670-5138}, P.S.~Hussain\cmsorcid{0000-0002-4825-5278}, M.~Jeitler\cmsAuthorMark{2}\cmsorcid{0000-0002-5141-9560}, N.~Krammer\cmsorcid{0000-0002-0548-0985}, A.~Li\cmsorcid{0000-0002-4547-116X}, D.~Liko\cmsorcid{0000-0002-3380-473X}, M.~Matthewman, A.~Pfeiffer\cmsorcid{0000-0001-5328-448X}, J.~Schieck\cmsAuthorMark{2}\cmsorcid{0000-0002-1058-8093}, R.~Sch\"{o}fbeck\cmsAuthorMark{2}\cmsorcid{0000-0002-2332-8784}, M.~Shooshtari\cmsorcid{0009-0004-8882-4887}, M.~Sonawane\cmsorcid{0000-0003-0510-7010}, N.~Van~Den~Bossche\cmsorcid{0000-0003-2973-4991}, W.~Waltenberger\cmsorcid{0000-0002-6215-7228}, C.E.~Wulz\cmsAuthorMark{2}\cmsorcid{0000-0001-9226-5812}
% ...
\par}
$^{1}$Also at Yerevan State University, Yerevan, Armenia\\
$^{2}$Also at Technische Universit\"{a}t Wien, Vienna, Austria\\
% ...
